\documentstyle[prd,aps,preprint,tighten,epsf]{revtex}
\title{Renormalization of Long-wavelength Solution of Einstein Equation}
\date{April 21, 1999}
\author{Yasusada Nambu\thanks{e-mail: nambu@allegro.phys.nagoya-u.ac.jp}}
\address{Department of Physics, Graduate School of Science, Nagoya University, Nagoya 464-8602, Japan}
\author{Yoshiyuki Y. Yamaguchi\thanks{e-mail: yyama@kuamp.kyoto-u.ac.jp}}
\address{Department of Applied Mathematics and Physics, Graduate School of Informatics, \\ Kyoto University, Kyoto 606-8501, Japan}
\begin{document}
%\tightenlines
\draft
\maketitle
%%%% DPNU number %%%%%%%%%%%%%%%%%%%%%%%%%
\vspace{-8 cm}
%\vspace{-4 cm}
\begin{flushright}
 DPNU-99-09
\end{flushright}
\vspace{8 cm}
%\vspace{4 cm}
%%%%%%%%%%%%%%%%
\begin{abstract}
 Using the renormalization group method, we improved the first order solution of the long-wavelength expansion of the Einstein equation. By assuming that the renormalization group transformation has the property of Lie group, we can regularize the secular divergence caused by the spatial gradient terms and absorb it to the background seed metric. The solution of the renormalization group equation shows that the renormalized metric  describes the behavior of gravitational collapse in the expanding universe qualitatively well.
\end{abstract}

%%%% local macro %%%%%%%%%%%%%%%%%%%%%%%%
\def\pa{\partial}
\def\ep{\epsilon}
%%%%%%%%%%%%%%%%%%%%%%%%%%%%%%%%%%%%%%%%%%%%%%%%%%%%%%%%%%%%%%%
\section{introduction}
%%%%%%%%%%%%%%%%%%%%%%%%%%%%%%%%%%%%%%%%%%%%%%%%%%%%%%%%%%%%%%%
The  spatial gradient expansion \cite{ge:a,ge:b,ge:c,ge:d} of the Einstein equation is non-linear approximation   method which describes long-wavelength  inhomogeneity in the universe. This approximation scheme expands the Einstein equation by the order of the spatial gradient. As the background solution, we solve the Einstein equation by neglecting all spatial gradient terms. The resultant solution has the same form as that for the spatially flat Friedman-Robertson-Walker(FRW) universe, but the three metric can have  spatial dependence. It is possible to include the effect of spatial gradient terms by calculating the next order. This method can describes the long-wavelength non-linear perturbation without imposing any symmetry for a space, and is suitable for analyzing the global structure of an inhomogeneous universe. Furthermore, it is easy to include perfect fluid with pressure and scalar fields.

But this scheme is valid only for the perturbation whose wavelength is larger than the Hubble horizon scale. For the matter field which satisfies energy conditions, the perturbation terms induced by the spatial gradient terms grows in time and finally dominates the background solution. This occurs when the wavelength of the perturbation equals the Hubble horizon scale. After this time, the wavelength of the perturbation becomes shorter than the horizon  and the result of the gradient expansion becomes unreliable. 

The similar situation occurs in the field of non-linear dynamical systems. To obtain temporal evolution of the solution of a non-linear differential equation, we usually apply perturbative expansion. But naive perturbation often yields secular terms due to the resonance phenomena. The secular terms prevent us from getting approximate but global solutions. There are many techniques to circumvent the problem, for example, averaging method, multi-time scale method, WKB method and so on\cite{appro:a,appro:b}. Although these method yield globally valid solutions, they provide no systematic procedure  for general dynamical systems because we must select a suitable assumption on the structure of perturbation series.

Renormalization group method\cite{rg:a,rg:b,rg:c,rg:d,rg:e} as a tool for global asymptotic analysis of the solution to differential equations unifies the techniques listed above, and can treat many systems irrespective of their features. Starting from a naive perturbative expansion, the secular divergence is absorbed to the constants of integration contained in the zeroth order solution by the renormalization procedure. The renormalized constants obey the renormalization group equation. This 
 method can be viewed as a tool of system reduction.  The renormalization group equation corresponds to the amplitude equation which describes slow motion dynamics in the original system. We can describe complicated dynamics contained in the original equation by extracting a simpler representation using the renormalization group method. 

In this paper, we apply the renormalization group method to the gradient expansion of the Einstein equation. Our purpose is to  obtain the renormalized  long-wavelength solution of the Einstein equation which is also valid for late time. Through the procedure of renormalization, we extract slow motion from the Einstein equation.

This paper is organized as follows. In section 2, we briefly review the gradient expansion. In section 3, we introduce the renormalization group method by using two examples. In section 4, the renormalization group method is applied to the solution of the gradient expansion. In section 5, we solve the renormalization group equation for several situations. Section 6 is devoted to summary and discussion. We use the unit in which $c=\hbar=8\pi G=1$ throughout the paper.
%%%%%%%%%%%%%%%%%%%%%%%%%%%%%%%%%%%%%%%%%%%%%%%%%%%%%%%%%%%%%%%
\section{long-wavelength expansion of Einstein+dust system}
%%%%%%%%%%%%%%%%%%%%%%%%%%%%%%%%%%%%%%%%%%%%%%%%%%%%%%%%%%%%%%%
We use synchronous reference frame:
%%%
\begin{equation}
  ds^2=-dt^2+\gamma_{ij}(t,x)dx^idx^j.
\end{equation}
%%%
The Einstein equation with dust fluid is 
%%%%%%%%%%%%%%
\begin{eqnarray}
 {}^{(3)}R&+&K^2-K^k_lK^l_k=2\rho, \label{eq:hc}\\
 K_{,i}&-&K^j_{i;j}=\sqrt{1+u^ku_k}\,\rho\,u_i, \label{eq:mc}\\
{}^{(3)}R^j_i&+&\frac{1}{\sqrt{\gamma}}\frac{\pa}{\pa t}\left(\sqrt{\gamma}\,K^j_i\right)=\frac{\rho}{2}\left(\delta^j_i+2u^ju_i\right),\label{eq:evo}\\
K_{ij}&=&\frac{1}{2}\frac{\pa\gamma_{ij}}{\pa t},
\end{eqnarray}
%%%%%%%%%%%%%
where $K_{ij}$ is the extrinsic curvature, $\rho$ is the energy density of dust, $u_i$ is the three velocity of dust. If we neglect terms with spatial derivative, the solution can be written
%%%%%%%%%
\begin{equation}
 \gamma^{(0)}_{ij}=a^2(t)h_{ij}(x),\quad \rho^{(0)}=\rho_0(t),\quad u_i^{(0)}=0,
\end{equation}
%%%%%%%%%%%%%%
where $h_{ij}(x)$ is an arbitrary function of spatial  coordinate (seed metric), and the scale factor $a(t)$ and the energy density $\rho_0(t)$ obey usual spatially flat FRW equation. Using this as the zeroth order solution, we solve this system by the following gradient expansion:
%%%%%%%%%%%%%%%%
\begin{eqnarray}
 \gamma_{ij}&=&a^2(t)h_{ij}(x)+\left(f_2(t)R_{ij}(h)+g_2(t)R(h)h_{ij}\right)+\cdots,\\
u_i&=&u_3(t)\nabla_iR(h)+\cdots,\\
\rho&=&\rho_0(t)+\rho_2(t)R(h)+\cdots,
\end{eqnarray}
%%%%%%%%%%%%%%%%
where $R_{ij}(h)$ is the Ricci tensor of the seed metric $h_{ij}$.
 The zeroth order solution is 
%%%%%%%%%%%%%%%%%
\begin{equation}
 \gamma^{(0)}_{ij}=t^{4/3}h_{ij}(x),\quad u^{(0)}_i=0,\quad \rho^{(0)}=\frac{4}{3t^2}.
\end{equation}
The solution to the  next order is
%%%%%%%%%%%%%%
\begin{eqnarray}
 \gamma^{(2)}_{ij}&=&-\frac{9t^2}{5}\left(R_{ij}(h)-\frac{1}{4}R(h)h_{ij}\right), \\
 \rho^{(2)}&=&\frac{3t^{-4/3}}{10}R(h),\quad u^{(3)}_i=0. \nonumber
\end{eqnarray}
%%%%%%%%%%%%%%%
The solution up to the second order of spatial gradient can be written as follows:
\begin{eqnarray}
\gamma_{ij}&=&t^{4/3}\left\{h_{ij}-\frac{9}{5}t^{2/3}\left(R_{ij}(h)-\frac{1}{4}R(h)h_{ij}\right)\right\},\\
\rho&=&\frac{4}{3t^2}\left(1-\frac{9}{20}t^{2/3}R(h)\right),\quad u_i=0.\nonumber
\end{eqnarray}
%%%%%%%%%%%%%%
We can see that the perturbation term,  which originated from the spatial gradient of the seed metric, grows as the universe expands and finally has the same amplitude as the background term at $t\sim H^{-1}$ when the wavelength of the perturbation equals Hubble horizon scale. After this time, the wavelength of perturbation becomes smaller than horizon scale and the long-wavelength expansion breaks down.
 To  make the gradient expansion applicable to the perturbation whose wavelength is smaller than the horizon scale, we use the renormalization group method.
%%%%%%%%%%%%%%%%%%%%%%%%%%%%%%%%%%%%%%%%%%%%%%%%%%%%%%%%%%%%%%%
\section{renormalization group method}
%%%%%%%%%%%%%%%%%%%%%%%%%%%%%%%%%%%%%%%%%%%%%%%%%%%%%%%%%%%%%%%
The renormalization group method\cite{rg:a,rg:b,rg:c,rg:d,rg:e} improve the long time behavior of naive perturbative expansion. We explain the basic concept of the renormalization group method using two examples. First one is a  harmonic oscillator. The equation of motion is
%%%%%%%%%%%%%%%%
\begin{equation}
 \ddot x+x=-\ep\, x,
\end{equation}
%%%%%%%%%%%%%%%
where $\ep$ is a small parameter. We solve this equation perturbatively by expanding the solution with respect to $\ep$:
%%%%%%
\begin{equation}
 x=x_0+\ep\,x_1+\cdots.
\end{equation}
%%%%%%
The solution up to $O(\ep)$ becomes
%%%%%%
\begin{equation}
 x=B_0\cos t+C_0\sin t+\frac{\ep}{2}\,(t-t_0)(C_0\cos t-B_0\sin t)+O(\ep^2),\label{eq:naive}
\end{equation}
%%%%%
where $B_0$ and $C_0$ are constants of integration determined by the initial condition at arbitrary time $t=t_0$. This naive perturbation breaks down when $\ep\,(t-t_0)>1$ because of the secular term. To regularize the perturbation series, we introduce an arbitrary time $\mu$, split $t-t_0$ as $t-\mu+\mu-t_0$, and absorb the divergent term  containing $\mu-t_0$ into the renormalized counterparts $B$ and $C$ of $B_0$ and $C_0$, respectively.

We introduce renormalized constants as follows:
%%%%%%%%%%%%%%%%
\begin{equation}
B_0=B(\mu)+\ep\,\delta B(\mu,t_0),\quad C_0=C(\mu)+\ep\,\delta C(\mu,t_0), \label{eq:counter}
\end{equation}
%%%%%%%%%%%%%%%
where $\delta B$ and $\delta C$ are counter terms that absorb the terms containing $\mu-t_0$ in naive solution. Substitute Eq.(\ref{eq:counter}) to Eq.(\ref{eq:naive}), we have
%%%%%%%%%%%
\begin{eqnarray}
 x&=&B(\mu)\cos t+C(\mu)\sin t \nonumber \\
 &&\quad +\ep \left\{\delta B\cos t+\delta C\sin t+\frac{1}{2}(t-\mu+\mu-t_0)(C(\mu)\cos t-B(\mu)\sin t)\right\}.
\end{eqnarray}
%%%%%%%%%%%%
$\delta B$ and $\delta C$ are chosen as
\begin{equation}
 \delta B(\mu,t_0)+\frac{1}{2}(\mu-t_0)C(\mu)=0,\quad \delta C(\mu,t_0)-\frac{1}{2}(\mu-t_0)B(\mu)=0. \label{eq:count}
\end{equation}
%%%%%%%%%%%%%
Using the relation $\ep\,\delta B=B_0-B(\mu),\, \ep\,\delta C=C_0-C(\mu)$,
\begin{equation}
 B(t_0)=B(\mu)-\frac{\ep}{2}(\mu-t_0)C(\mu),\quad C(t_0)=C(\mu)+\frac{\ep}{2}(\mu-t_0)B(\mu). \label{eq:rgtransf}
\end{equation}
%%%%%%%%%%%%%%%%
These equations define the transformation up to $O(\ep)$:
$${\cal R}_{\mu-t_0}:~(B(t_0), C(t_0))\rightarrow (B(\mu), C(\mu)).$$
%%%%%%%%%%%%%
Explicit form of the transformation is
%%%%
\begin{equation}
\left(
  \begin{array}{c}
     B(\mu) \\ C(\mu)
  \end{array}
\right)={\cal R}_{\mu-t_0}
\left(
   \begin{array}{c}
     B(t_0) \\ C(t_0)
  \end{array}
\right)=
\left(
   \begin{array}{cc}
    1 & \frac{\ep}{2}(\mu-t_0) \\
   -\frac{\ep}{2}(\mu-t_0) & 1
   \end{array}
\right)
\left(
   \begin{array}{c}
     B(t_0) \\ C(t_0)
  \end{array}
\right)+O(\ep^2).
\end{equation}
%%%%%%%%%%%%%%
${\cal R}$ satisfies the composition law:
%%%%%%%%%%%%%
\begin{eqnarray}
{\cal R}_{\mu-\mu_1}\otimes{\cal R}_{\mu_1-t_0}&=&
\left(
  \begin{array}{cc}
   1 & \frac{\ep}{2}(\mu-\mu_1) \\
 -\frac{\ep}{2}(\mu-\mu_1) & 1
  \end{array}
\right)
\left(
  \begin{array}{cc}
   1 & \frac{\ep}{2}(\mu_1-t_0) \\
 -\frac{\ep}{2}(\mu_1-t_0) & 1
  \end{array}
\right)+O(\ep^2) \nonumber \\
&=&
\left(
  \begin{array}{cc}
   1 & \frac{\ep}{2}(\mu-t_0) \\
 -\frac{\ep}{2}(\mu-t_0) & 1
  \end{array}
\right)+O(\ep^2)\nonumber \\
&=&{\cal R}_{\mu-t_0},
\end{eqnarray}
%%%%%%%%%%%%%
and ${\cal R}_0=1$ and ${\cal R}_{-\mu}$ is the inverse transformation of ${\cal R}_\mu$. So we can conclude that the transformation forms a Lie group up to $O(\ep)$. Assuming the properties of Lie group, we can extend the locally valid expression (\ref{eq:rgtransf}) to a global one, which is valid for arbitrary large $\mu-t_0$. We  apply this transformation to get $(B(\mu),C(\mu))$ at arbitrary large $\mu$. By differentiating Eq.\,(\ref{eq:rgtransf}) with respect to $\mu$ and setting $t_0=\mu$, we have renormalization group equations:
%%%%%
\begin{equation}
 \frac{\pa B}{\pa\mu}=\frac{\ep}{2}\,C(\mu),\quad \frac{\pa C}{\pa \mu}=-\frac{\ep}{2}\,B(\mu). \label{eq:rge}
\end{equation}
%%%%%%
The renormalized solution becomes
\begin{equation}
 x(t)=B(\mu)\cos t+C(\mu)\sin t+\frac{\ep}{2}\,(t-\mu)(C(\mu)\cos t-B(\mu)\sin t). \label{eq:rsola}
\end{equation}
%%%%%%%%%%%%%%%%
Solving the renormalization group equation (\ref{eq:rge}) and equating $\mu$ and $t$ in (\ref{eq:rsola}) eliminates the secular term and we get uniformly valid result:
\begin{equation}
 x(t)=B(0)\cos\left(1+\frac{\ep}{2}\right)t+C(0)\sin\left(1+\frac{\ep}{2}\right)t.
\end{equation}
%%%%%%%%%%%%%%%%%%%%%%%%%%%%%%%%%%%%%%%%%%%%%%%%%
%FRW case
%%%%%%%%%%

Second example is the Einstein equation for the FRW universe with dust. The spatial component of the Einstein equation is
%%%%%
\begin{equation}
 \ddot\alpha+\frac{3}{2}\left(\dot\alpha\right)^2=-\frac{\ep}{2}e^{-2\alpha}, \label{eq:frw}
\end{equation}
%%%%%%
where $\alpha(t)$ is the logarithm of the scale factor of the universe $a(t)$ and $\ep$ is the sign of the spatial curvature. The exact solution is given by
%%%%%%
\begin{equation}
 a(t)=e^{\alpha(t)}=\left\{
  \begin{array}{lll}
    a_0\,(1-\cos\eta), & t=a_0\,(\eta-\sin\eta) & \quad\mbox{for}~\ep=1 \\
    a_0\,\frac{\eta^2}{2}, & t=a_0\,\frac{\eta^3}{6} & \quad\mbox{for}~\ep=0 \\
    a_0\,(\cosh\eta-1), & t=a_0\,(\sinh\eta-\eta) & \quad\mbox{for}~\ep=-1
  \end{array}
  \right. \label{eq:esol}
\end{equation}
%%%%%%%
We solve Eq.\,(\ref{eq:frw}) perturbatively by assuming that the right hand side is small. This is expansion with respect to small spatial curvature around the flat universe. By substituting $\alpha=\alpha_0+\ep\,\alpha_1+\cdots$ to Eq.\,(\ref{eq:frw}), we have naive solution:
%%%%%
\begin{equation}
  \alpha=\ln\tau+C_0-\ep\,\frac{9}{20}e^{-2C_0}(\tau-\tau_0)+O(\ep^2), \label{eq:nsol}
\end{equation}
%%%%%
where we introduced a new time variable $\tau=t^{2/3}$ and $C_0$ is a constant of integration determined by the initial condition at $\tau=\tau_0$. $O(\ep)$ term is secular and we regularize this term by introducing arbitrary time $\mu$ and renormalized constant $C_0=C(\mu)+\ep\,\delta C(\mu,\tau_0)$:
%%%%
\begin{equation}
 \alpha=\ln\tau+C(\mu)+\ep\,\delta C(\mu,\tau_0)-\ep\,\frac{9}{20}e^{-2C(\mu)}(\tau-\mu+\mu-\tau_0).
\end{equation}
%%%%
The counter term $\delta C$ is determined by so as to absorb $\mu-\tau_0$ dependent term:
%%%%
\begin{equation}
 \delta C(\mu,\tau_0)-\frac{9}{20}e^{-2C(\mu)}(\mu-\tau_0)=0.
\end{equation}
%%%%
This defines the renormalization group transformation ${\cal R}_{\mu-\tau_0}:~C(\tau_0)\ \rightarrow\ C(\mu)$
%%%%
\begin{equation}
 C(\mu)=C(\tau_0)-\ep\,\frac{9}{20}e^{-2C(\mu)}(\mu-\tau_0),
\end{equation}
%%%%%
and this transformation forms Lie group up to $O(\ep)$. So we can have $C(\mu)$ for arbitrary large value of $\mu-\tau_0$ by assuming the property of Lie group, and this makes possible to produce globally uniform approximated solution of the original equation. The renormalization group equation is
%%%
\begin{equation}
  \frac{\pa C(\mu)}{\pa\mu}=-\ep\,\frac{9}{20}e^{-2C(\mu)},
\end{equation}
%%%
and the solution is
%%%
\begin{equation}
  C(\mu)=\frac{1}{2}\ln\left(c-\frac{9\ep}{10}\mu\right),
\end{equation}
%%%
where $c$ is a constant of integration. The renormalized scale factor is given by
%%%
\begin{equation}
  a(t)=e^{\alpha(t)}=\tau\,e^{C(\tau)}=t^{2/3}\left(c-\frac{9\ep}{10}t^{2/3}\right)^{1/2}. \label{eq:rsol}
\end{equation}
%%%%

As the zeroth order solution, it is possible to include another integration constant $t_0$ which defines the origin of cosmic time $t$. By requiring that the renormalization group transformation forms Lie group, it can be shown that $t_0$ does not get renormalization. So it is sufficient to consider the solution with the boundary condition $a(t=0)=0$ which fixes the value of $t_0$ equals zero. This point is different from the example of harmonic oscillator, in which 
 case two integration constants $B$ and  $C$ both get renormalization.

We compare the renormalized solution (\ref{eq:rsol}) with the exact solution (\ref{eq:esol}) and the naive solution (\ref{eq:nsol}) for the case of a  closed universe ($\ep=1$). The scale factor of the exact solution has maximum at $t=\pi$ and goes to zero at $t=2\pi$. The naive solution does not show this behavior. The renormalized solution improves the naive solution and reproduces the expanding and contracting feature of the exact solution (Fig.1).
%%%%%%%% figure %%%%%%%%
%\begin{figure}[hbtp]
%\label{fig}
%\centering
%  \epsfbox{fig1.eps}
%  \caption[fig]{The evolution of the scale factor for a closed FRW universe with dust. The solid curve is the exact solution, thin dashed curve is the naive solution and thick dashed curve is the  renormalized solution.}
%\end{figure}

%%%%%%%%%%%%%%%%%%%%%%%%%%%%%%%%%%%%%%%%%%%%%%%%%%%%%%%%%%%%%%%
\section{renormalization of long-wavelength solution}
%%%%%%%%%%%%%%%%%%%%%%%%%%%%%%%%%%%%%%%%%%%%%%%%%%%%%%%%%%%%%%%
Now we proceed to the long-wavelength solution of Einstein equation. We renormalize the secular behavior of three metric $\gamma_{ij}$. By introducing a new time variable $\tau=t^{2/3}$, the longwavelength solution can be written as
%%%%
\begin{equation}
 \gamma_{ij}=\tau\,\left\{h_{ij}(x)-\frac{9\tau}{5}\left(R_{ij}(h)-\frac{1}{4}R(h)h_{ij}\right)\right\}.
\end{equation}
%%%%%
We renormalize the secular term $\frac{9\tau}{5}(R_{ij}-\frac{1}{4}R h_{ij})$. We introduce the initial time $\tau_0$ by re-defining the seed metric $h_{ij}$, and define renormalized metric and the counter term  $h_{ij}(x)=h_{ij}(x,\mu)+\delta h_{ij}(x,\mu,\tau_0)$:
%%%%%
\begin{eqnarray}
 h_{ij}(x)&-&\frac{9}{5}(\tau-\tau_0)\left(R_{ij}-\frac{1}{4}R(h)h_{ij}\right) \nonumber \\
 &=&h_{ij}(x,\mu)+\delta h_{ij}-\frac{9}{5}(\tau-\mu+\mu-\tau_0)\left(R_{ij}-\frac{1}{4}R(h)h_{ij}\right).
\end{eqnarray}
%%%%%
The counter term is determined so that it absorbs terms containing $\mu-\tau_0$:
\begin{equation}
 \delta h_{ij}(\mu,\tau)-\frac{9}{5}(\mu-\tau_0)\left(R_{ij}(h)-\frac{1}{4}R(h)h_{ij}\right)=0.
\end{equation}
%%%%%%%%%%
Using the definition of $\delta h_{ij}$,
\begin{equation}
 h_{ij}(x,\mu)=h_{ij}(x,\tau_0)-\frac{9}{5}(\mu-\tau_0)\left(R_{ij}(h(\mu))-\frac{1}{4}R(h(\mu))h_{ij}(\mu)\right). \label{eq:rgtrnsf2}
\end{equation}
%%%%%%%%%%%%
This equation defines the renormalization group transformation ${\cal R}_{\mu-\tau_0}:h_{ij}(\tau_0)\rightarrow h_{ij}(\mu)$ and this transformation is Lie group up to $O(\ep)$. We therefore can get the value of $h_{ij}(\mu)$ for arbitrary $\mu$ using the relation (\ref{eq:rgtrnsf2}) by assuming the property of Lie group. The renormalization group equation is obtained by differentiating Eq.\,(\ref{eq:rgtrnsf2}) with respect to $\mu$ and setting $\tau_0=\mu$:
%%%
\begin{equation}
 \frac{\pa}{\pa\tau}h_{ij}(x,\mu)=-\frac{9}{5}\left[R_{ij}(h(\mu))-\frac{1}{4}R(h(\mu))h_{ij}(\mu)\right], \label{eq:rge2}
\end{equation}
%%%%%%%%%%%%
and renormalized solution is
\begin{eqnarray}
 \gamma_{ij}&=&t^{4/3}h_{ij}(x,t), \\
 \rho &=&\frac{4}{3t^2}+\frac{3t^{-4/3}}{10}R(h(x,t)).
\end{eqnarray}
%%%%%%%%%%%%%%%%%%%%%%%%%%%%%%%%%%%%%%%%%%%%%%%%%%%%%%%%%%%%%%%
\section{solution of renormalization group equation}
%%%%%%%%%%%%%%%%%%%%%%%%%%%%%%%%%%%%%%%%%%%%%%%%%%%%%%%%%%%%%%%
We solve the renormalization group equation (\ref{eq:rge2}) for some special cases and see how the renormalization group method improves the behavior of the long-wavelength solution.
%%%%%%%%%%%%%%%%%%%%%%
\subsection{FRW case}
The metric is
$$
 h_{ij}(x,t)=\Omega^2(t)\,\sigma_{ij}(x),\quad R_{ij}(\sigma)=\frac{1}{3}\sigma_{ij}\,R(\sigma),
$$
where $\sigma_{ij}(x)$ is the metric of three dimensional maximally symmetric space. In this case, the renormalization group equation reduces to
$$
 \frac{\pa}{\pa\tau}\Omega^2(\tau)=-\frac{9}{10}k,\quad (k=\pm1, 0),
$$
and the solution is
$$
 \Omega(t)=\sqrt{c-\frac{9k}{10}t^{2/3}},
$$
where $c$ is a constant of integration. 
The renormalized solution is
%%%%%%%%%%%%%%%%
\begin{eqnarray}
 \gamma_{ij}&=&t^{4/3}\left(c-\frac{9k}{10}\,t^{2/3}\right)\sigma_{ij}(x), \nonumber \\
 \rho&=&\frac{4}{3t^2}\left(1+\frac{27}{20}\frac{k\,t^{2/3}}{c-\frac{9k}{10}\,t^{2/3}}\right), \label{sol:frw}
\end{eqnarray}
%%%%%%%%%%%%%%
and the scale factor of the universe is given by
\begin{equation}
 a(t)=t^{2/3}\sqrt{c-\frac{9k}{10}t^{2/3}}.
\end{equation}
%%%%%%%%%%%%%%%
This is the same as the solution (\ref{eq:rsol}).

%%%%%%%%%%%%%%%%%%%%%%%%%%%%%%%%%%%%%%%%%%%%%%%%%%%%%%%%%%%%%%%%%%%%%%%%%%
\subsection{spherically symmetric case}
In spherical coordinate $(r, \theta, \phi)$, the metric is 
\begin{equation}
 h_{ij}=\mbox{diag}\left(A^2(\tau, r), B^2(\tau, r), B^2(\tau, r)\sin^2\phi\right).
\end{equation}
%%%%%%%%%%%%%%%%%
The renormalization group equation (\ref{eq:rge2}) becomes
\begin{eqnarray}
 2A\frac{\pa A}{\pa\tau}&=&\frac{9}{5}\left(\frac{A^2}{2B^2}-\frac{A_{,r}B_{,r}}{AB}-\frac{\left(B_{,r}\right)^2}{2B^2}+\frac{B_{,rr}}{B}\right), \quad (rr~ \mbox{component})\\
 2B\frac{\pa B}{\pa\tau}&=&\frac{9}{5}\left(-\frac{1}{2}+\frac{\left(B_{,r}\right)^2}{2A^2}\right).\quad (\theta\theta~ \mbox{component})
\end{eqnarray}
%%%%%%%%%%%%%%%
The solution of this equation is given by
%%%%%%%%%%%
\begin{eqnarray}
 B&=&\left(1-\frac{9\alpha(r)}{10}\tau\right)^{1/2}\beta(r), \\
 A&=&\frac{B_{,r}}{\sqrt{1-\alpha(r)\,\beta^2(r)}},
\end{eqnarray}
%%%%%%%%%%%%%%
where $\alpha(r)$ and $\beta(r)$ are  arbitrary function of $r$. The renormalized metric and density are
%%%%%%
\begin{eqnarray}
  ds^2&=&-dt^2+\frac{\left(t^{2/3}B_{,r}\right)^2}{1-\alpha\,\beta^2}\,dr^2+\left(t^{2/3}B\right)^2d\Omega^2_2,\label{eq:rgstb} \\
  \rho&=& \frac{4}{3t^2}+\frac{3}{5t^{4/3}}\frac{3\alpha\,\beta_{,r}\left(1-\frac{9\alpha}{10}\tau\right)+\alpha_{,r}\,\beta\,\left(1-\frac{27\alpha}{20}\tau\right)}{\left(1-\frac{9\alpha}{10}\tau\right)\left[\left(1-\frac{9\alpha}{10}\tau\right)\beta_{,r}-\frac{9}{20}\alpha_{,r}\,\beta\,\tau\right]}.
\end{eqnarray}
%%%%%%%%%%%%%%
By choosing $\beta=r, \alpha=0,\pm 1$, we can recover the solution of FRW case (\ref{sol:frw}). This solution corresponds to Toleman-Bondi solution\cite{TB}:
%%%%%%
\begin{equation}
  ds^2=-dt^2+\frac{R_{,r}(t,r)^2}{1+2E(r)}dr^2+R(t,r)^2 d\Omega_2^2
\end{equation}
%%%%%%
where
%%%
\begin{equation}
R(t,r)=\left\{
 \begin{array}{lll}
 -\frac{M(r)}{2E(r)}(1-\cos\eta),& t=\frac{M(r)}{(-2E(r))^{3/2}}(\eta-\sin\eta) & \quad\mbox{for}~E(r)<0 \\
   \left(\frac{9M(r)}{2}\right)^{1/3}t^{2/3}, & &\quad\mbox{for}~E(r)=0 \\
 \frac{M(r)}{2E(r)}(\cosh\eta-1), & t=\frac{M(r)}{(2E(r))^{3/2}}(\sinh\eta-\eta) &\quad\mbox {for}~E(r)>0 
 \end{array}
 \right.
\end{equation}
%%%%%%
and
%%%%%%%%%%%
\begin{equation}
 \rho(t,r)=\frac{2M_{,r}(r)}{R^2\,R_{,r}}.
\end{equation}
%%%%%%%%%%%%
$E(r), M(r)$ are arbitrary function of radial coordinate $r$ and are connected to the initial distribution of the spatial curvature and  the initial distribution of mass density of dust, respectively. They have the following correspondence with the functions $\alpha, \beta$ contained in the renormalization group solution (\ref{eq:rgstb}):
%%%%%%
\begin{equation}
   2E(r)\leftrightarrow -\alpha(r)\beta^2(r),\qquad \frac{9}{2}M(r)\leftrightarrow \beta^3(r).
\end{equation}
%%%%%%
The renormalized solution well reproduces the feature of the metric of spherically symmetric gravitational collapse of dust.
%%%%%%%%%%%%%%%%%%%%%%%%%%%%%%%%%
\subsection{Szekeres solution}
In Cartesian coordinate $(x,y,z)$, the metric is assumed to be
\begin{equation}
 h_{ij}=\mbox{diag}\left(1, 1, A^2(\tau,x,y,z)\right).
\end{equation}
%%%%%%%%%%%%%%
The renormalization group equation (\ref{eq:rge2}) reduces to the following three equations:
%%%%%%%%%%%%%%%%
\begin{eqnarray}
 0&=&-\frac{A_{,xy}}{A}, \quad (xy~\mbox{component})\nonumber \\
 0&=&\frac{A_{,yy}-A_{,xx}}{2A}, \quad (xx~\mbox{component})\\
 \frac{\pa A^2}{\pa\tau}&=&-\frac{1}{2}A(A_{,xx}+A_{,yy}), \quad(zz~\mbox{component})\nonumber
\end{eqnarray}
%%%%%%%%%%%%%%%%%%%
The solution of this equation is
\begin{equation}
 A=g(z)(x^2+y^2)+\frac{9}{5}g(z)\,\tau+c(z),
\end{equation}
%%%%%%%%%%%%%
where $c, g$ are arbitrary function of $z$. The renormalized metric is 
\begin{equation}
 ds^2=-dt^2+t^{4/3}\left[dx^2+dy^2+\left(g(z)\,(x^2+y^2)+\frac{9g(z)}{5}t^{2/3}+c(z)\right)^2\,dz^2\right].
\end{equation}
%%%%%%%%%%%%%%
This is the Szekeres' exact solution\cite{seke}, which represents one dimensional gravitational collapse. It is known that the ``naive"  gradient expansion reproduces this solution by including the fourth order spatial gradient\cite{ge:c}. We obtained the solution using the second order spatial gradient with renormalization. In this  case, renormalization procedure much improves the naive solution.

%%%%%%%%%%%%%%%%%%%%%%%%%%%%%%%%%%%%%%%%%%%%%%%%%%%%%%%%%%%%%%%
\section{summary and discussion}
%%%%%%%%%%%%%%%%%%%%%%%%%%%%%%%%%%%%%%%%%%%%%%%%%%%%%%%%%%%%%%%
 We have investigated renormalization of the long-wavelength solution of the Einstein equation which has secular behavior. We applied the renormalization group method to improve the long time behavior of the solution. We introduced renormalization point  $\mu\sim L/H^{-1}$ where $L$ is physical wavelength  of perturbation. For $\mu\ll 1(L\gg H^{-1})$, the naive expansion(gradient expansion) well approximate the exact solution. But as the universe expands,  $\mu\sim H^{-1}/L\propto t^{1/3}$ becomes large and at $\mu\sim 1(L\sim H^{-1})$, perturbative expansion breaks down. After this time, the wavelength of the perturbation is smaller than the horizon scale. Using renormalization transformation which has the property of Lie group, we can renormalize the secular terms  caused by the spatial gradient of the background seed metric. The renormalized metric can be extended to large value of $\mu$, which corresponds to the small scale fluctuation. 

We obtained the solution of the renormalization group equation for FRW, spherically symmetric and Szekeres cases. The behavior of the renormalized solution indicates that they  describe the collapsing phase of the system qualitatively well. The renormalization group method is regarded as the procedure of system reduction. This means the renormalization group Eq.\,(\ref{eq:rge2}) is reduced version of the original Einstein equation and describes slow motion dynamics of the original equation.  We expect the renormalization group equation (\ref{eq:rge2}) has physically interesting properties and solutions which contained in the Einstein equation.

We can look the renormalization of the long-wavelength solution from the view point of the backreaction problem in cosmology. The naive solution represents the evolution of the perturbation with  the fixed background metric. By renormalizing the naive solution, the constants $h_{ij}(x)$ contained  in the background solution  becomes to have the time dependence due to the spatial inhomogeneity. So we can investigate how the spatial inhomogeneity affects ``background'' metric $h_{ij}$ by solving the renormalization group equation. The remarkable feature of the  renormalization group approach is that it does not need any spatial averaging which is necessary in conventional approach to the backreaction problem in cosmology\cite{back}. 

 In this paper, we considered only dust fluid as the matter field. But the gradient expansion can treat the perfect fluid with pressure and scalar fields. For the scalar field system, although we cannot write down  exact solutions of the naive expansion in general, it is possible to define the renormalization group transformation and obtain the renormalized solution formally. It is interesting to apply the renormalization group method to the inflationary model and investigate the effect of the stochastic quantum noise on the background geometry. We can construct the inhomogeneous model of stochastic inflation\cite{inf:a,inf:b,inf:c,inf:d} and this is our next subject to be  explored.
%%%%%%%% ack %%%%%%%%%%%%%%%%%
\vspace{2cm}
\begin{center}
{\bf ACKNOWLEDGEMENT}
\end{center}

We would like to thank K. Nozaki for indicating us the possibility of higher order renormalizability of the long-wavelength solution.
%%%%%%%%%%%%%%%%%%%%%%%%%%%%%%%%%%%%%%%%%%%%%%%%%%%%%%%%%%%%%%%%

%%%%%%%%%%%%%%%%%%%%%%% figure %%%%%%%%%%%%%%%%%%%%%%%%%%%%%%%%%%%%%%%%%%%
\begin{figure}[hbtp]
\label{fig}
\centering
  \epsfbox{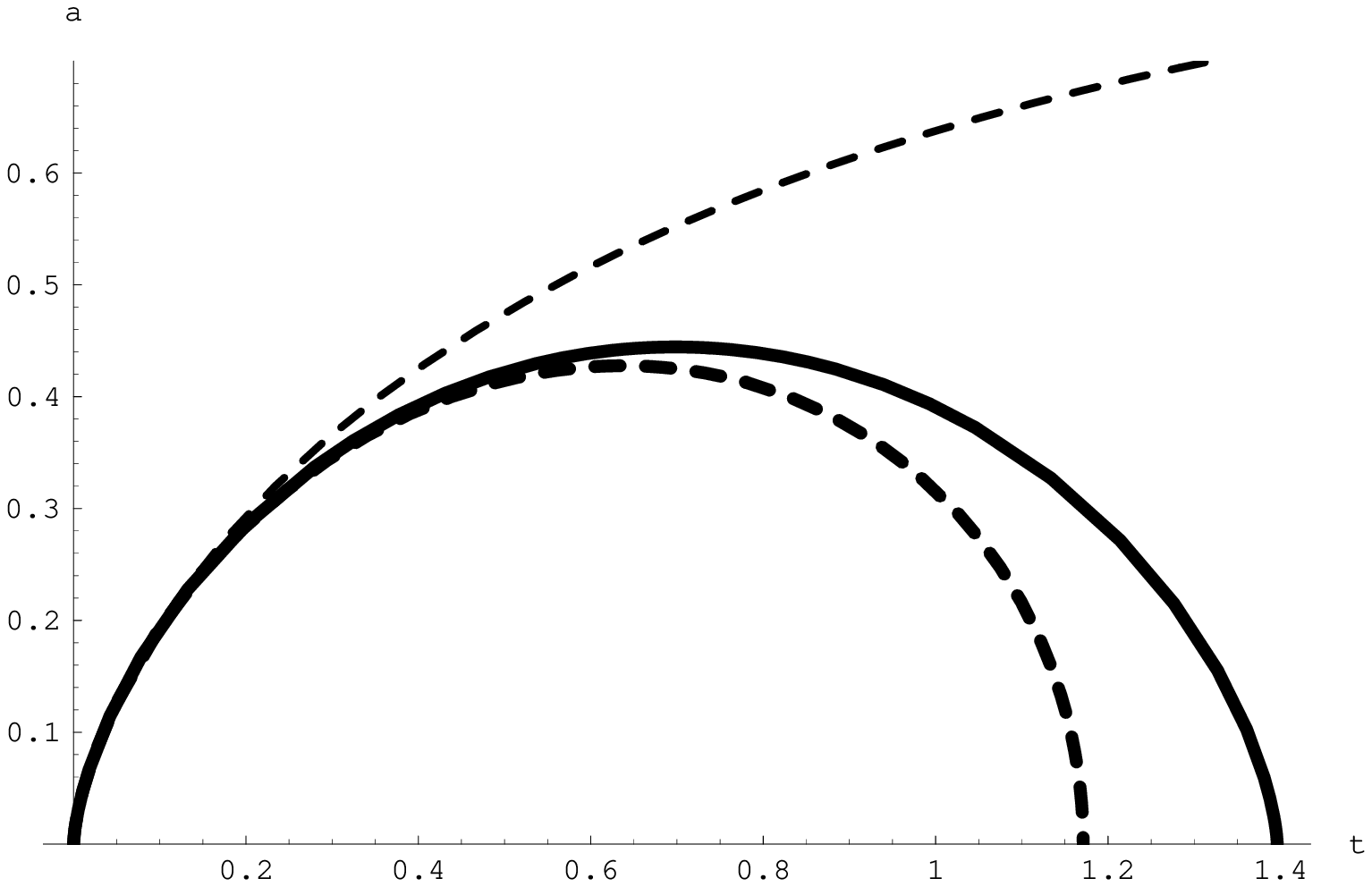}
  \caption[fig]{The evolution of the scale factor for a closed FRW universe with dust. The solid curve is the exact solution, thin dashed curve is the naive solution and thick dashed curve is the  renormalized solution.}
\end{figure}

\end{document}